\documentclass[aps,pre,twocolumn,showpacs,groupedaddress]{revtex4}
\usepackage{graphicx}
\usepackage{bm}
\usepackage{amsmath, amsthm, amssymb}
\usepackage{eurosym}
\usepackage{tabularx}

\begin{document}

\title{Imitation versus payoff - duality of the decision-making process
demonstrates criticality and consensus formation}
\author{M. Turalska$^{1}$}
\author{B.J. West$^{1,2}$}
\affiliation{
$^{1}$ Physics Department, Duke University, Durham, NC 27709, USA\\
$^{2}$ Information Science Directorate, US Army Research Office, Research
Triangle Park, NC 27708, USA
}
\date{\today }

\begin{abstract}
We consider a dual model of decision making, in which an individual forms its opinion based on contrasting mechanisms of imitation and rational calculation. The decision making model (DMM) implements imitating behavior by means of a network of coupled two-state master equations that undergoes a phase transition at a critical value of a control parameter. The evolutionary spatial game (EGM), being a generalization of the Prisoner's dilemma game, is used to determine in objective fashion the cooperative or anti-cooperative strategy adopted by individuals. Interactions between two sources of dynamics increases the domain of initial states attracted to phase transition dynamics beyond that of the DMM network in isolation. Additionally, on average the influence of the DMM on the game increases the
final observed fraction of cooperators in the system.
\end{abstract}

\pacs{ 02.50.Le, 89.75.Da, 02.50.Ey, 05.50.+q }
\maketitle

\section{Introduction}

In a society interconnected by family ties, friendships, acquaintances or work relations it is unavoidable that a person's behavior or decisions depends on the choices made by other people. The surrounding social network
influences the opinions we hold, the products we buy or the activities we pursue. Therefore exploring the basic principles that give rise to social processes in which individual behavior aggregates into collective outcomes
can provide significant insight into the individual's decision-making process.

Human performance with regard to decision-making can be viewed from at least two complementary perspectives. One is from the assumptions made about how individuals behave and to predict the outcomes of their interactions at a
group level. This outlines a mechanistic approach to the study of decision-making. The predictions based on this approach are tested by comparing them with observational data of how individuals behave in response to other individuals and to the environment. In this way the mechanisms through which collective behavior is generated are determined. At the turn of the twentieth century Tarde \cite{tarde98} argued that imitation was the fundamental mechanism by which the phenomena of crowds, fads, fashions and crime, as well as other collective behaviors, could be understood. At the same time, Baldwin \cite{baldwin97} maintained that the behavior based on imitation arose out of the mental development of the child resulting from imitation being a basic form of learning.

Currently, imitation remains an important concept in the social sciences, being pointed to as a mechanism responsible for herding, information cascades \cite{banerjee92} or many homophily-based behaviors \cite{christakis09}. Social experiments such as the Friends and Family study \cite{petland2011} demonstrated higher efficiency of incentives directed at the social network of an individual rather than directly offered to a person, suggesting strong influence that the actions of our peers have on our own decisions \cite{petland2013}. Similarly, coping proved to be the preferred and most effective strategy to acquire adaptive behavior in complex
environment \cite{randel2010}, even when other non-social sources of
information were available at the same cost.

This mechanistic approach to decision-making is separate and distinct from
the functional approach, in which we ask what is the value or function of a
particular behavioral strategy. The basic assumption of the second approach
is that a given behavior can be rationally evaluated in terms of costs and
benefits, which allows for an objective comparison of alternative
strategies. This principle of balancing costs against benefits to arrive at
a decision is central to modern economical \cite{becker}, political \cite%
{politics} and social science \cite{social}.

In mathematical terms the functional approach aligns with the basic
assumptions of game theory, which originated from games of chance. Game
theory influenced behavioral sciences through the introduction of the
utility function by Daniel Bernoulli in 1730 \cite{bernoulli30}. In doing so
Bernoulli resolved the famous St. Petersburg paradox \cite{shlesinger87},
demonstrating that a rational strategy should be based on the subjective
desirability of a game's outcome rather than being proportional to the
game's expected value. The suggestion that the value of a thing to an
individual is not simply equivalent to its monetary value reached its full
articulation in the voices of von Neumann and Morgenstern \cite{vonneumann53}
in their seminal work on game theory and economics.

More recently game theory was used to study the emergence of cooperative
behaviors, as a way of obtaining insight into this evolutionary puzzling
phenomena. The work of Nowak and May \cite{nowak92} for the first time
extended game theory principles to spatial networks, and demonstrated that
the introduction of spatial structure between players lead to spatially and
temporally rich dynamics. Following this observation, the impact of the
spatial structure on the evolution of cooperation has been investigated in
detail \cite{nowak93, huberman93, Killingback96, szabo1998, szabo2007}.
Contrary to the well-mixed case, where non-cooperative behavior is favored,
the well-known Prisoner's Dilemma game performed on a square lattice with
next-nearest-neighbor interactions promotes cooperation. In the effort to
investigate the impact of different interaction topologies, heterogenous
topologies were investigated more recently, with scale-free architecture
being the most extensively studied \cite{santos2005, santos2006}.

Herein we explore both mechanistic and functional perspectives as a way to
understand the decision-making process. We consider a society to be given by
a two-layer network, whose elements are individuals making decisions
simultaneously using two distinct sets of criteria, as indicated in Fig. \ref%
{fig_latticeN}. On the one hand, individuals form their decisions based on
the perception of actions and appearances of their neighbors, adopting the
concept of imitation. This behavior, captured by the decision making model
(DMM) \cite{turalska09,turalska11} demonstrates the cooperative behavior
induced by the critical dynamics associated with a phase transition. This
behavior is counterbalanced by the rationality of the evolutionary game
model (EGM). Thus, while personal decisions are influenced by the desire to
be liked and accepted, individuals also weigh the effect of certain
potential relations on their careers, balancing the cost and payoffs of such
relations. This latter behavior is captured by the rational and
deterministic EGM rules in the spirit of the original approach of Nowak and
May \cite{nowak92}. Thus, the adopted two-layer network model allows us to
refer to two aspects of the decision-making process of a single individual,
each aspect being defined by distinct dynamic rules. The two layers interact
with one another and modify their separate dynamics.

In Section \ref{DMM} we review the basic properties of the DMM network in
isolation, that is, not interacting with the second network. The phase
transition properties of the DMM network for different interaction strengths
are briefly discussed. In Section \ref{EGM} we outline the basic properties
of the EGM network in isolation. In this case the traditional game theory
final states are identified. Finally, in Section \ref{TLM} the two
functionally distinct networks are allowed to interact with one another and
the difference in the asymptotic states due to their mutual interaction are
analyzed and discussed. We draw some conclusions in Section \ref{conclusions}.

\begin{figure}[t]
\includegraphics{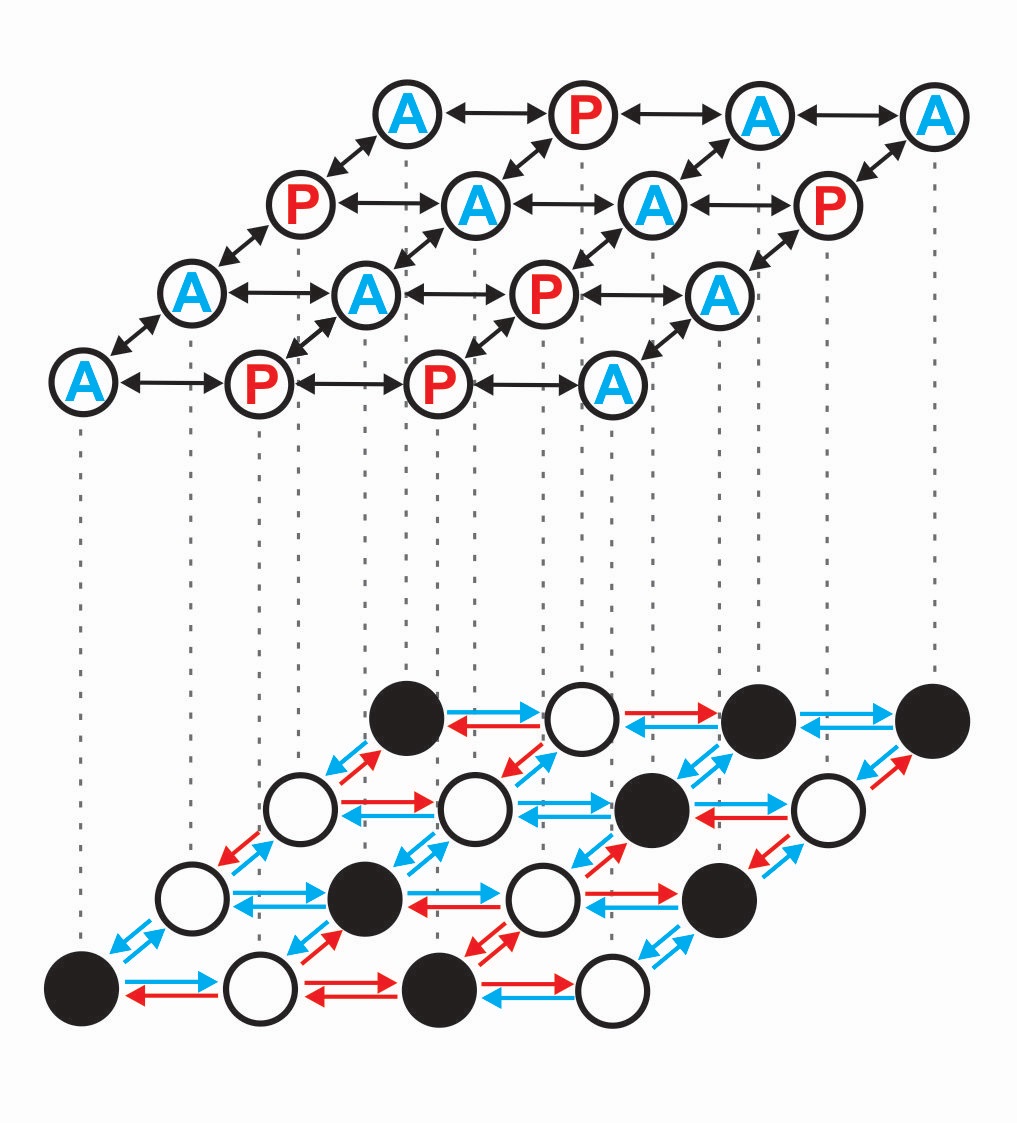}
\caption{ \emph{(Color online)} Decision-making process in a society presented as a two-layer network. The lower layer models the imitation behavior, with positive (+1) and negative (-1) opinions held by an individual of the decision making model being represented by black and white nodes, respectively. The upper layer models the rational behavior of the evolutionary game, with nodes selecting between two strategies: being a protagonist (P, blue circle) or antagonist (A, red circle). The color of the arrows in the lower layer (P, protagonist - red; A, antagonist - blue) indicates that strategy adopted in the upper layer influences the character of the imitating behavior.}
\label{fig_latticeN}
\end{figure}

\section{Isolated DMM Layer\label{DMM}}

In the DMM network the state of an isolated individual $s_{i}(t)$ is
described by a two-state master equation,
\begin{equation}
\frac{d\mathbf{p}^{\left( i\right) }(t)}{dt}=\mathbf{G}^{\left( i\right) }(t)%
\mathbf{p}^{\left( i\right) }(t),  \label{master}
\end{equation}%
where $\mathbf{G}^{\left( i\right) }(t)$ is a $2\times 2$ transition matrix:
\begin{equation}
\mathbf{G}^{\left( i\right) }(t)=\left[
\begin{array}{ll}
-g_{_{+-}}^{\left( i\right) }(t) & g_{_{-+}}^{\left( i\right) }(t) \\
-g_{_{-+}}^{\left( i\right) }(t) & g_{_{+-}}^{\left( i\right) }(t)%
\end{array}%
\right]  \label{coupling}
\end{equation}%
and the probability of being in one of two states $(+1,-1)$, is $\mathbf{p}%
^{\left( i\right) }(t)=(p_{+}^{\left( i\right) },p_{-}^{\left( i\right) })$.
Positioning $N$ such individuals at the nodes of a square lattice yields a
system of $N$ coupled two-state master equations \cite{turalska09,bianco08}
which, under the assumption of nearest neighbor interactions, contain
time-dependent transition rates for each of the $i$ individuals:
\begin{align}
g_{_{+-}}^{\left( i\right) }(t)& = & & g_{0}\exp \left[ \frac{K}{M}\left(
M_{+}(i,t)-M_{-}(i,t)\right) \right] ;  \notag \\
g_{_{-+}}^{\left( i\right) }(t)& = & & g_{0}\exp \left[ -\frac{K}{M}\left(
M_{+}(i,t)-M_{-}(i,t)\right) \right]  \label{tran}
\end{align}%
where $K$ is the strength of the interaction. On the two-dimensional lattice
$M=4$ and $0\leq M_{\pm }(i,t)\leq 4$ denotes the count of nearest neighbors
in states $\pm 1$ at time $t$. In the mean field approximation $M_{\pm
}(i,t)/M\rightarrow p_{\pm }^{\left( i\right) }\left( t\right) $ and the
transition rates become exponentially dependent on the state probabilities
resulting in a highly nonlinear master equation \cite{west14}.

\begin{figure}[t]
\includegraphics{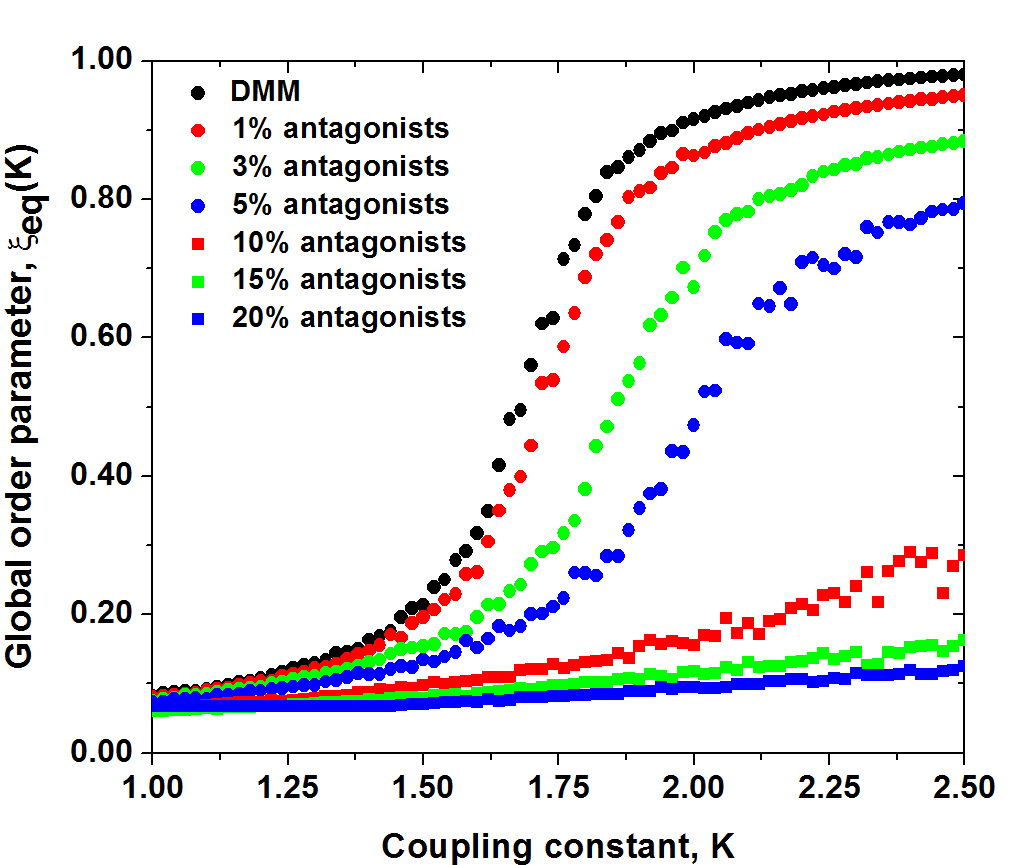}
\caption{ \emph{(Color online)} The phase transition property
of a predominantly protagonist DMM lattice is indicated with an increasing
number of antagonists. The black dots denote an all protagonist lattice, the
dots receeding from the first calculation depict 1\%, 3\%, 5\%, \ and the
squares depict 10\%, 15\% and 20\% antagonists in sequence.}
\label{fig_phaseN}
\end{figure}

A wealth of results exist for the dynamics of DMM on networks of various
topologies, in a configuration of a single network, as well as in the case
of coupled networks \cite{west14}. Here we concentrate on the global
behavior of the model, which is defined by the fluctuations of the global
variable
\begin{equation}
\xi \left( K,t\right) =\frac{1}{N}\underset{i=1}{\overset{N}{\sum }}%
s_{i}(K,t)  \label{global}
\end{equation}%
which is further used to calculate the equilibrium global value $\xi _{eq}$ $%
\equiv \left\langle \left\vert \xi \left( K,t\right) \right\vert
\right\rangle .$ When the coupling parameter $K>0$, single units of the
system become more and more cooperative and for coupling value larger than a
critical one, $K>K_{C}$, the interaction between units is strong enough to
give rise to a majority state, during which a significant number of nodes
adopts the same opinion at the same time. Thus, the global dynamics of the
DMM is characterized by a phase transition with respect to the coupling
parameter $K$ (Fig. \ref{fig_phaseN}), demonstrating that a system of
identical units imitating each other's actions is able to reach consensus,
given sufficient influence of the imitation on their decisions.

However, this model society in which all members are interacting only
through positive relationships (friendships, collaborations or sharing of
information) is not very realistic, since negative effects are also present
in most circumstances. Some relationships are friendly, while others are
antagonistic or even hostile and interactions between people often lead to
disagreement and conflict. Thus, there is a need to modify the basic DMM\ to
include a mix of positive and negative relationships.

Rather than define the nature of the relationship between two nodes, which
is usually done by introducing links with positive (reciprocal friendship)
and negative (mutual antagonism) signs \cite{kleinberg10}, we consider a
network in which the nature of the interaction between individuals depends
on the individuals themselves. Thus, our modeled society is composed of two
kinds of individuals: those that always cooperate (protagonists) and those
that always oppose the opinion of their peers (antagonists). As a result we
observe three kinds of interaction between two nodes: reciprocal friendship
and mutual antagonism, as well as cooperator-antagonist pairs, in which one
node wants to cooperate while another opposes any kind of mutual action.

On the one hand cooperating individuals operate according to the DMM
dynamics defined in Eq. \ref{tran}. On the other hand, antagonists at any
time oppose the opinion of their neighbors, and their dynamics are defined
by the transition rates
\begin{align}
g_{_{+-}}^{\left( i\right) }(t)& = & & g_{0}\exp \left[ -\frac{K}{M}\left(
M_{+}(i,t)-M_{-}(i,t)\right) \right] ;  \notag \\
g_{_{-+}}^{\left( i\right) }(t)& = & & g_{0}\exp \left[ \frac{K}{M}\left(
M_{+}(i,t)-M_{-}(i,t)\right) \right]   \label{antago}
\end{align}%
where the only difference with respect to the dynamics of protagonists is an opposite sign of the coupling constant.

All numerical calculations in this section are performed on a square lattice of $N=20$ $\times $ $20$ nodes with periodic boundary conditions. The initial state of each individual is randomly assigned. In a single time step
a computer calculation involving the entire lattice is performed and for every element $s_{i}$ the transition rate of either Eq. \ref{tran} or Eq. \ref{antago} is calculated, according to which element is given the possibility to change its state. The transition rate for a non-interacting unit is $g_{0}=0.01$. The equilibrium value $\xi _{eq}$ is calculated as an average over $10^{6}$ consecutive time steps, after the same number of time steps since the initialization has passed thereby insuring that all transient behavior has died away. The assignment of protagonist and antagonist behavior is done randomly.

The dynamics of the DMM lattice with increasing numbers of antagonists is depicted on Figure \ref{fig_phaseN}. It is evident from the figure that the phase transition the DMM dynamics undergoes is sensitive to the fraction of
the network members that are antagonists. The DMM lattice dynamics undergoes a phase transition and this criticality persists with up to 5\% antagonists randomly placed on the lattice. However, the phase transition is clearly
suppressed when the number of antagonists is 10\% and above. These results are consistent with those determining the influence of committed minorities on group consensus \cite{xie11,turalska13}.

\section{Isolated EGM layer\label{EGM}}

The evolutionary game model (EGM) used herein is a generalization of the
Prisoners' Dilemma (PD) game. Traditionally the PD game consists of two
players, each of whom may choose to cooperate or defect in any single
encounter. If both players choose to cooperate, both get a reward pay-off of
size $R$; if one defects and the other cooperates the defector receives a
"temptation" pay-off $T$ while the cooperator receives a "sucker's" pay-off $%
S$; if both defect, both receive "punishment" pay-off $P$. \ The ordering of
the payoff parameters given by $T>R>P>S$ defines the PD game. Nowak and May
\cite{nowak92} considered a two-dimensional lattice over which the PD game
was played in a sequential fashion, where at each time step every node was
able to change its strategy (defect or cooperate) depending on the outcome
of the game played with its neighbors at the previous time step. The
historical nomenclature of cooperator and defector is here replaced with
protagonist and antagonist, respectively, which we believe to be more
compatible with the language of the two-layer network model.

\begin{figure}[t]
\includegraphics[scale=2.00]{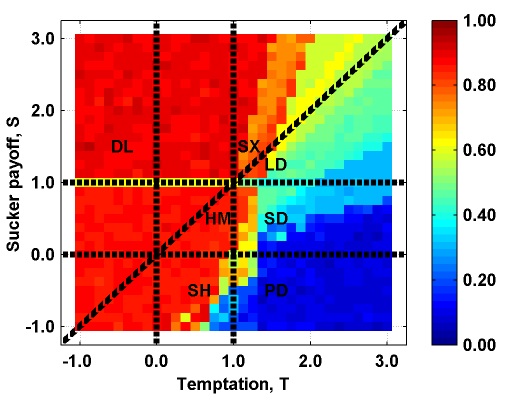}
\caption{ \emph{(Color online)} The equilibrium fraction of protagonists present in the EGM lattice of $N=20$ x $20$ nodes is plotted for the game parameters $R=1$ and $P=0$. The initial fraction of protagonists was $50\%$, randomly distributed over the lattice. Periodic boundary conditions are considered. The region $T>1,S\leq 0$ locates the Prisoner's Dilemma (PD) game; the Stag Hunt (SH) game is found in the domain $0\leq T\leq1,S\leq 0;$ the Snowdrift (SD) game is bounded by $T\geq 1,0\leq S\leq 1;$ the Leader (LD) game is confined by $T=1$ and to the
right of the $S=T$ diagonal; and the Battle of the Sexes (SX) is above the
diagonal $S=T$ and bordered on the left by $T=1$.}
\label{fig_prisonerN}
\end{figure}

Consider the EGM network dynamics without it being coupled to the DMM network dynamics. As depicted on the top part of Figure \ref{fig_latticeN}, the protagonists and antagonists are placed on the sites of a two-dimensional lattice and interact only with their four nearest neighbors. In each generation every individual plays a deterministic game defined by a pay-off matrix $\left(
\begin{array}{cc}
R & S \\
T & P%
\end{array}%
\right) $ with all its neighbors. The pay-off gained by each individual at the end of each generation is determined by summing payoffs of $2\times 2$ games with each of its neighbors. The scores in the neighborhood, including
the individual's own score are ranked, and in the following generation the individual adopts the strategy of the most successful player from among its neighbors. In the case of a tie between the scores of cooperative and antagonistic players, the individual keeps its original strategy. Thus, adopted evolutionary strategy is to act like the most successful neighbor.

Even this simple and completely deterministic situation leads to a wide
array of behaviors. Figure \ref{fig_prisonerN} depicts the equilibrium
fraction of protagonists present in the EGM game as a function of the two
parameters $S$ and $T$. Without loss of generality, we assume that $R>P$ and
normalize the pay-off values such that $R=1$ and $P=0$. The initial
configuration of the game consists of 50\% protagonists and antagonists,
distributed randomly on the lattice.

It is evident that in the domain $T\leq 1,$ for almost all values of $S,$
that the equilibrium state is dominated by protagonists. These regions are
determined to have between 5\% and 10\% randomly distributed antagonists
asymptotically. Whereas for $T\geq 1$ and $S<0,$ the region of the PD game,
antagonists dominate with the network having between 5\% and 10\% randomly
distributed protagonists. The remaining regions have differing levels of
protagonists at equilibrium. The traditional games for which there is a
substantial literature are marked and are not addressed here in more detail.
We merely note that selected $ST$ parameter values enable us to determine
the outcome for the two-layer network and thereby determine the mutual
influence of the DMM and EGM dynamics and the relative influences of
imitation and payoff on decision making in all these cases.

\section{Two-layer Network\label{TLM}}

The constant struggle between maximizing individual gains and the desire to
be part of a community is modeled by an interaction between EGM and DMM
layers (Fig. \ref{fig_latticeN}). The coupling is realized dynamically,
since the behavior of a node in the DMM layer depends on its strategy in the
EGM layer, being protagonist or antagonist. In return the local
configuration of nodes in the DMM layer can change the strategy of a node in
the EGM layer.

More precisely, each time step of the simulation consists of four operations:

\begin{enumerate}
\item One step of the EGM dynamics is realized in the EGM layer. The pay-off
for each individual is evaluated and used to update its strategy in the next
generation.

\item The following generation of protagonists and antagonists is used to
define the sign of a coupling constant $K$ in the DMM layer. Consequently, a
node in the DMM layer may be a protagonist at one time step, while in the
next it acts as an antagonist, due to the changes made to the EGM layer.

\item One step of the DMM dynamics is performed in the DMM layer, allowing
nodes to change their state from $\pm 1$ to $\mp 1$, or not change at all.
This step potentially affects the local neighborhood of an individual, when
at one time step an individual is surrounded by mostly other individuals of
the same sign, and in the next time step it is surrounded by individuals of
opposite sign.

\item A change to the DMM layer finally affects the strategy of an
individual in the EGM layer. The decision to change strategy in the EGM
layer is made if the average state of the local neighborhood of an
individual in the DMM layer is of an opposite sign with respect to that
individual.
\end{enumerate}

After those four steps, one full iteration of the two-layer network is
completed, and the time index is advanced.

\begin{figure}[t]
\includegraphics[scale=0.25]{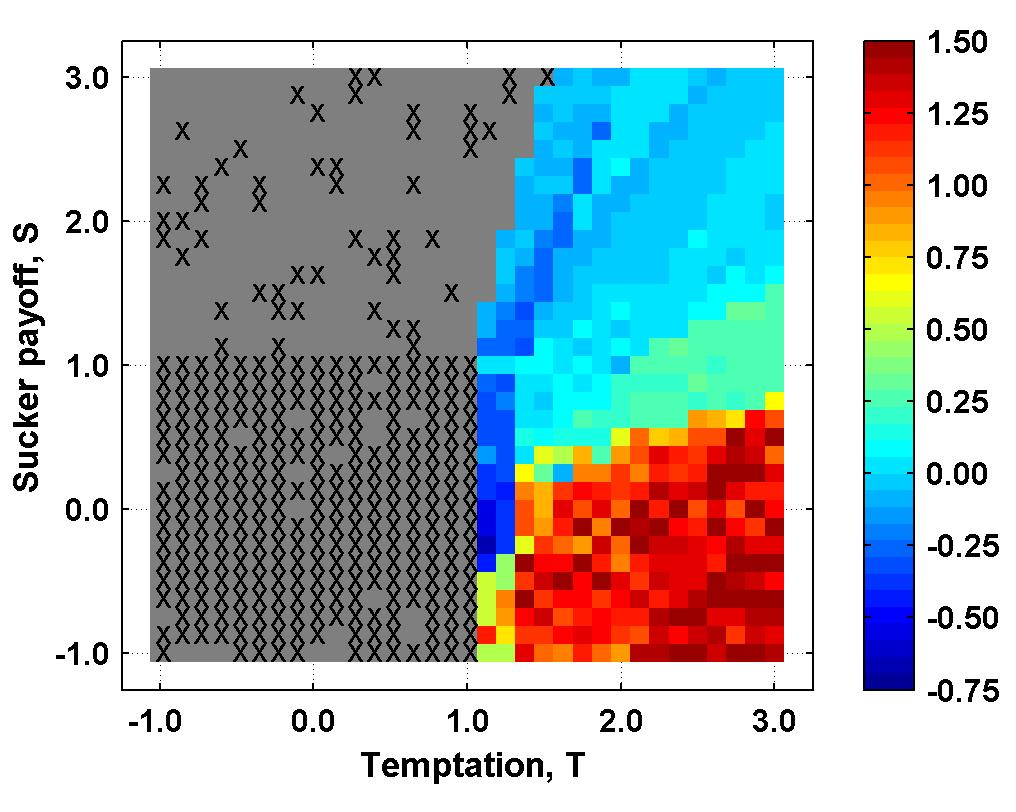}
\caption{ \emph{(Color online)} The ratio $D$ of the final fraction of
protagonists observed in the DMM-EGM system to the fraction of protagonists
observed in the EGM network in isolation is color coded. Grey area depicts the range of parameter values whose DMM-EGM network dynamics give rise to phase transition. \textbf{X} depicts $ST$
values for which fraction of protagonists in the isolated EGM network is
smaller then $90\%$.}
\label{fig_differenceN}
\end{figure}

Figure \ref{fig_differenceN} summarizes the asymptotic dynamics of the
two-layer network for a broad set of $S$ and $T$ values. The color at each
value of $S$ and $T$ is determined as the ratio of the final fraction of
protagonists observed in the DMM-EGM network to the fraction of protagonists
observed in the EGM network in isolation:
\begin{equation}
D=\log _{10}\left[ \frac{\text{Protagonists}_{DMM-EGM}}{\text{Protagonists}%
_{EGM}}\right]  \label{value}
\end{equation}

The large grey area depicts the range of parameter values whose DMM-EGM
network dynamics give rise to phase transition. The initial fraction of 50\%
protagonists increases with the phase transition to a final state of
complete consensus among protagonists. The symbol \textbf{X} depicts $ST$
values for which the fraction of protagonists in the isolated EGM network is
smaller then $90\%$ and consequently if the DMM were operating in isolation
on such a configuration of protagonists and antagonists, the phase
transition would be strongly suppressed. However the coupling of DMM-EGM
networks facilitates the critical transition at these values.

\begin{figure}[t]
\includegraphics{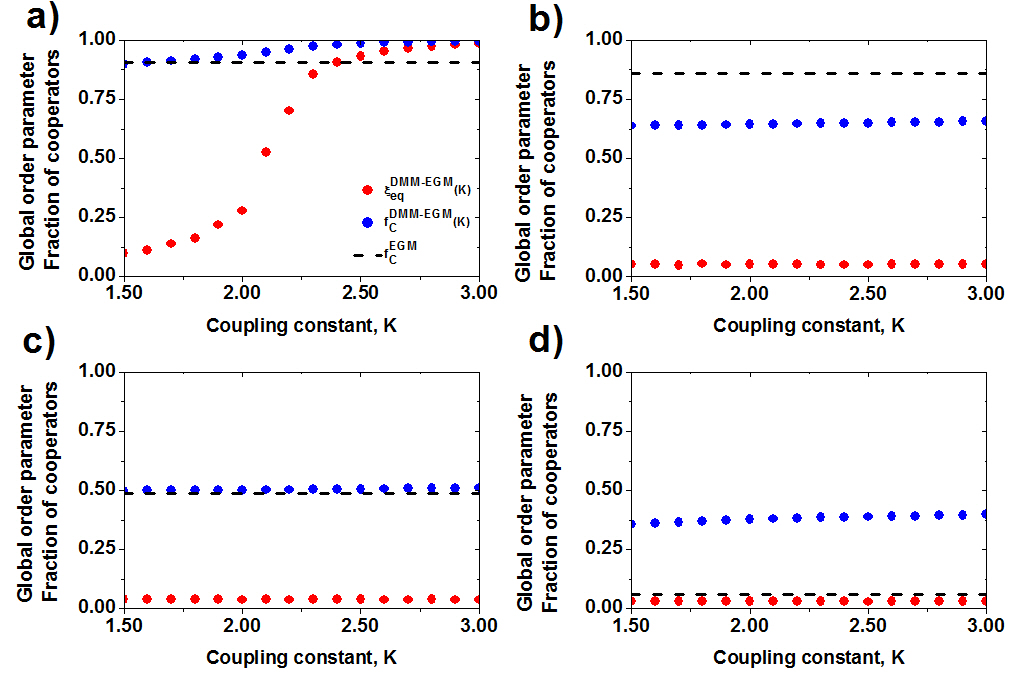}
\caption{ \emph{(Color online)} Typical equilibrium dynamical
behavior for the global variable is depicted in four regions of (S,T)-space.
The blue dots denote the equilibrium fraction of protagonists and the red dots denote value of the global variable in the two-layer system. The dashed line marks the equilibrium solution of the EGM in isolation. a) $T\leq 1,-1\leq S\leq 3$ global variable undergoes a phase transition at a $K_{c}\approx 2.2;$ b) $T\geq 1, S\leq 1$ (narrow band of Fig.\ref{fig_differenceN} with no phase transition), the level of protagonists is below that of EGM in isolation; c) the predominantly green region in Fig. \protect\ref{fig_prisonerN} with no phase transition and the fraction of protagonists the same as for the isolated EGM; d) the PD segment of Fig. \protect\ref{fig_prisonerN}\ has no phase transition and a low fraction of protagonists that is above the isolated EGM level.}
\label{fig_four}
\end{figure}

The remaining regions do not undergo phase transition, but do show a wide
range of dynamic behaviors. Four types of typical dynamics for the two-layer
DMM-EGM network can be determined. For each of the panels in Figure \ref%
{fig_four} the DMM-EGM network was initialized with 50\% antagonists
randomly distributed. Such an initial state is sufficient to suppress a
phase transition for the DMM in isolation (Fig. \ref{fig_phaseN}).

However in Figure \ref{fig_four}a, as the control parameter $K$ is
increased, the average global variable of the DMM layer in the DMM-EGM
network manifests critical behavior. Simultaneously, the 10\% antagonists
seen in isolated EGM layer change their strategy and are converted to
protagonists in the DMM-EGM network. For reference the dashed curve shows
the fraction of protagonists in the isolated EGM network.

\begin{figure}[t]
\includegraphics{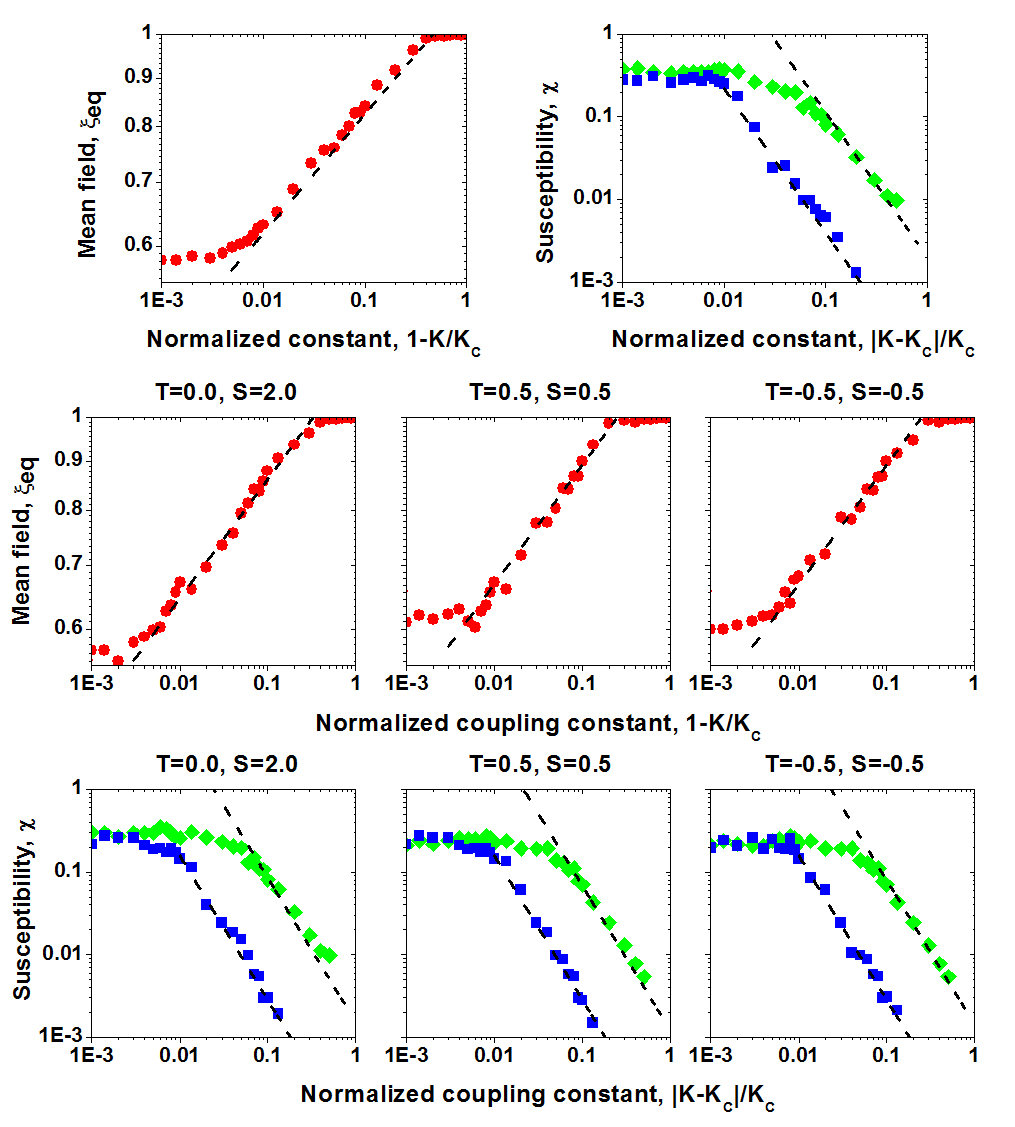}
\caption{ \emph{(Color online)} Scaling of the global order
parameter (top left panel, middle row) and the susceptibility (top right
panel, bottom row) in the vicinity of the phase transition point. Top panel
illustrates the scaling present in an isolated DMM\ layer with critical
control parameter $K_{C}=1.70$. In middle and bottom rows, the columns
correspond to the dual-layer configuration with $T=0,S=2$, $T=0.5,S=0.5$ and $%
T=-0.5,S=-0.5$ for the left, middle and right column respectively. Lattice
size is $N=20$ x $20$, and scaling exponents are $\protect\beta =1/8$ and $%
\protect\gamma =-7/4.$ }
\label{fig_scaling}
\end{figure}

We demonstrated in our earlier work \cite{turalska13} that the global
dynamics of the DMM belongs to the Ising universality class. The
investigation of the scaling properties of the global order parameter and
the susceptibility confirms that property also for the dual-layer dynamics.
Despite the relatively small size of the system, a lattice of $N=20$ x $20$
nodes, on which the DMM dynamics is realized in the isolated layer
configuration, the values of the mean field and susceptibility in the
vicinity of the phase transition point scale as a power law (Fig. \ref%
{fig_scaling} top row). When coupled to the EGM layer, the scaling behavior
and exponents are preserved (Fig. \ref{fig_scaling}), suggesting that a very
small fraction of antagonists present in the system does not cause a
significant change in dynamical properties of the system.

Figure \ref{fig_four}c depicts representative behavior of the DMM-EGM
dynamics for $ST$ values from the upper right quadrant of Fig. \ref%
{fig_prisonerN}: $S,T>1$. Here the fraction of protagonists is typically the
same as that of the isolated EGM (the pale blue region of Fig. \ref%
{fig_differenceN}). Figure \ref{fig_four}b is taken from a narrow transition
channel parallel to $T=1$ in which the fraction of protagonists in the dual
network is typically below that of the isolated EGM network, indicating a
marked change in behavior for small increases in temptation. Across this
channel the arrangement of the payoff matrix converts the behavior of the
social group from being cooperative and consensus seeking to being
antagonistic and disagreeable. Once in this region of $T\geq 1$ and $S\leq
0, $ the domain of the PD game, the fraction of protagonists can be greater
than that of the isolated EGM (Fig. \ref{fig_four}d) and still not produce a
phase transition and therefore the group does not reach consensus.

\begin{figure*}[ht]
\includegraphics{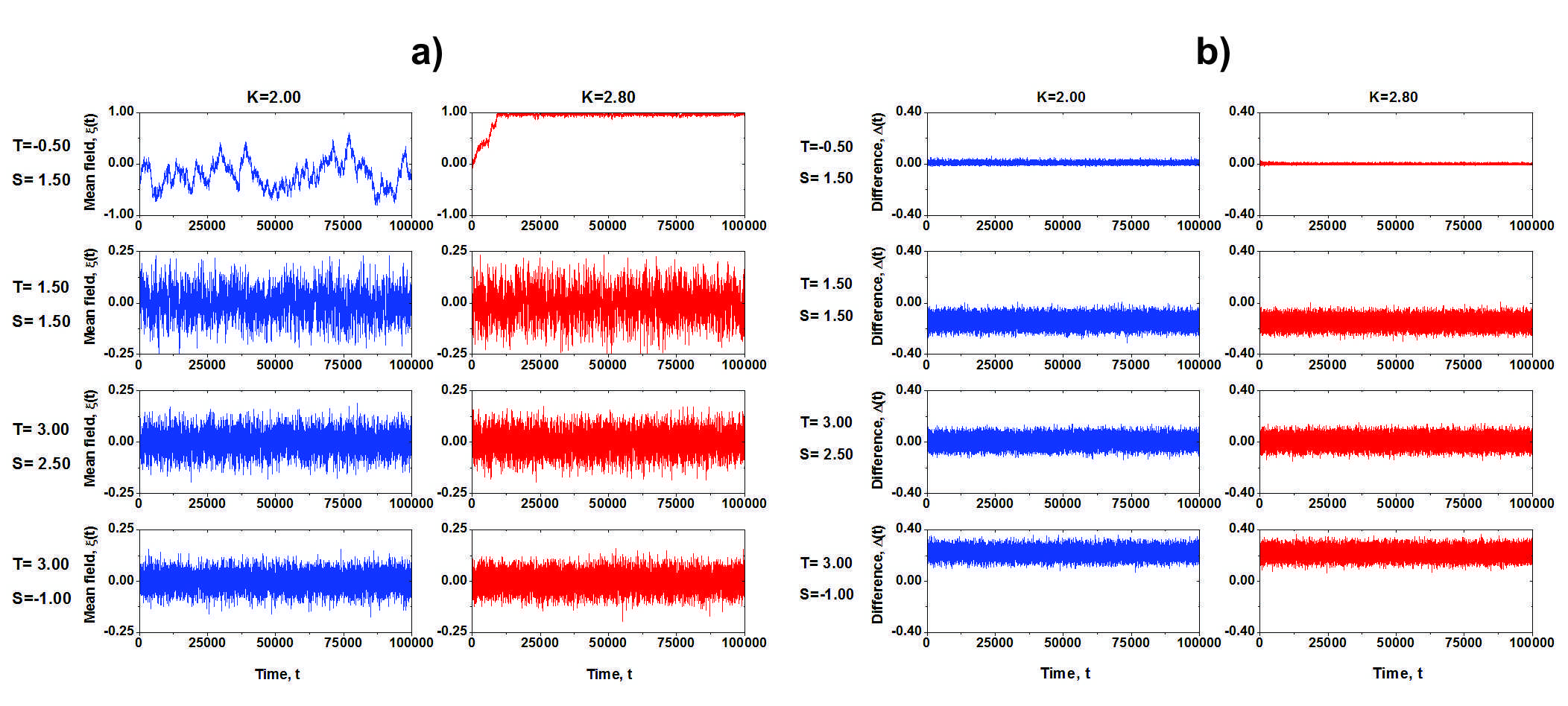}
\caption{ \emph{(Color online)} a) Here the global variable $\protect%
\xi \left( K,t\right) $ is depicted as a function of time for two values of $K$ and a sequence of $ST$ values. b) Plotted is the difference between the fraction of nodes whose strategy was changed from the EGM to the DMM layer minus that changed from the DMM to the EGM layer at each time of the simulation. }
\label{fig_time series}
\end{figure*}

The DMM-EGM network dynamics is depicted using the global variable $\xi(K,t) $ for a subcritical and a supercritical value of the control parameter at selected values of the sucker payoff $S$ and temptation $T$ in Figure \ref{fig_time series}a. In the subcritical case the DMM-EGM network dynamics are seen to be random. The global variable $\xi (K,t)$ time series have large
scale fluctuations for $T<0$ and $S>0$, with much more rapid large amplitude
fluctuations in the positive temptation regions. In the supercritical region
the DMM-EGM network transitions to consensus for $T<0$ are characterized by
a very low amplitude fluctuations once consensus is achieved. However in the
other $ST$ regions the value of the control parameter seems to be irrelevant
and large amplitude fluctuations of $\xi (K,t)$ are observed.

Another way to view the dynamics of the coupled DMM-EGM network is to record
the difference in the number of changes inflicted by the DMM layer on the
EGM layer and the number of changes inflicted by the EGM layer on the DMM
layer at each time step. Thus, it is the difference in the number of changes
performed at the second and fourth step of the simulation. The time series
recorded in Figure \ref{fig_time series}b have the same parameter values as
those depicted in Figure \ref{fig_time series}a. First we notice the
insensitivity of the behavior of the time series to the $K$ value of the
dynamics. Next is the balance in the two kinds of change apparent for
negative temptation $T<0$ with relatively low amplitude fluctuations.
Whether there are more changes projected from the DMM or EGM layer to the
complementary layer varies with the $S$ and $T$ values. For example at $T=3.0
$ the two changes are equal for $S=2.5$ and DMM exceeds EGM changes at $%
S=1.0,$ with relatively large fluctuations in both cases. This suggests that
four regions of dynamics of DMM-EGM network, identified on Fig. \ref%
{fig_four} are characterized by the balance between the DMM and EGM layer in
the case that presents phase transition, and an imbalance between the two
mechanisms of decision-making in other cases.

\section{Conclusion\label{conclusions}}

We expect that the behavior present in the dual layer model is not limited
to the regular lattice configuration discussed herein. The DMM demonstrates
the cooperative behavior in a wide range of network topologies, where the
phase transition is observed for both random, small-world and scale-free
configurations \cite{turalskaDiss}. Similarly, the cooperation based on the
Prisoner's Dilemma is present in heterogenous topologies \cite{gomes2007}.
Additionally, recent experimental studies of the cooperation when humans
play a Prisoner's Dilemma demonstrate that both the regular lattice and a
scale-free network reach the same level of cooperation, which is comparable
with the level of cooperation of smaller homogeneous networks \cite%
{Moreno2012, Lazaro2012}.

In this paper we considered two dynamically coupled networks; the dynamics
of one being determined by imitation using the DMM \cite{west14} and the
other following the game theoretic format prescribed by Nowak and May \cite%
{nowak92}. We find that in the domain $T\leq 1,R=1,P=0$ and all $S$ the
protagonists asymptotically dominate and phase transition behavior is
robust. Critical dynamics occurs in this region even when the initial
fraction of antagonists would be sufficient to inhibit critical behavior for
the DMM network in isolation. Consequently the EGM layer increases the
domain of initial states attracted to critical dynamics for the two-layer
network dynamics.

On the other hand, the influence of the EGM antagonists on the DMM-EGM
network dynamics is quite different from that of committed minorities \cite%
{xie11,turalska13}, even when the committed minorities are modeled as a
distinct dynamic network \cite{west14}. The EGM layer can actually increase
the stability of the consensus-making process for the two-layer network
model. Consequently, consensus can be facilitated by payoffs even in cases
where intuition might dictate that antagonists would prevail. This suggests
the possibility of counter-intuitive policies, which society might adopt,
that could facilitate the consensus-making of large groups, even in the face
of what might appear to be overwhelming polarization.

\section{Acknowledgement}

The authors would like to thank the Army Research Office for support of this
research.

\end{document}